\documentclass[%
 amsmath,amssymb,
 reprint,%
]{revtex4-1}

\usepackage{amsmath}
\usepackage[dvips]{graphicx}
\usepackage{epsfig}
\usepackage{tabularx}
\usepackage{color}
\usepackage[pdfpagemode=UseNone,colorlinks=true,linkcolor=blue,citecolor=blue]{hyperref}

\usepackage{soul}
\usepackage{gensymb}

\begin{document}

\title[]{Quadratic optomechanical cooling of a cavity-levitated nanosphere}

\author{N. P. Bullier}%
\affiliation{Department of Physics and Astronomy, University College London, Gower Street, London WC1E 6BT, United Kingdom}%

\author{A. Pontin}
\email{a.pontin@ucl.ac.uk}
\affiliation{Department of Physics and Astronomy, University College London, Gower Street, London WC1E 6BT, United Kingdom}

\author{P. F. Barker}
 \email{p.barker@ucl.ac.uk}
\affiliation{Department of Physics and Astronomy, University College London, Gower Street, London WC1E 6BT, United Kingdom}%

\begin{abstract}
We report on cooling the center-of-mass motion of a nanoparticle due to a purely quadratic coupling between its motion and the optical field of a high finesse cavity. The resulting interaction gives rise to a Van der Pol nonlinear damping, which is analogous to conventional parametric feedback where the cavity provides passive feedback without measurement. We show experimentally that like feedback cooling the resulting energy distribution is strongly nonthermal and can be controlled by the nonlinear damping of the cavity. As quadratic coupling has a prominent role in proposed protocols to generate deeply nonclassical states, our work represents a first step for producing such states in a levitated system.
\end{abstract}

\maketitle

The field of cavity optomechanics has made significant progress over last decade by controlling and tailoring the interaction between an optical field and a mechanical oscillator. Mechanical modes have been brought into the quantum regime\,\cite{Chan2011,Delic2020science}. This includes quantum mechanical squeezing of an oscillator below its zero point fluctuations\,\cite{Wollman952} and entanglement between two mechanical oscillators\,\cite{entanglement1,entanglement2}. More recently, the creation of oscillators levitated in vacuum using optical, electrical or magnetic fields have been realised. These systems offer greater decoupling from the environment, and they typically only have a few mechanical degrees of freedom\,\cite{peter2010,Chang1005,Romero_Isart_2010}. They also introduce new optomechanical degrees of freedom via their rotational motion\,\cite{Arita2013}.

By trapping charged nanospheres in a Paul trap, or in a focused laser beam, $\sim$nHz mechanical linewidths are theoretically achievable\,\cite{Chang1005}. Those large Q factors provide in theory long lifetimes for studying nonclassical states of motion. Of particular interest are nonlinearities in the optomechanical interaction\,\cite{quadratic7,quadratic10,quadratic11,quadratic12,quadratic2}. For example, the control over the position of ions or levitated nanoparticles within an optical cavity enables tuning between linear and quadratic optomechanical coupling\,\cite{quadratic1,giacomo2016,Delic2019,Delic2020}. In membranes, single-photon to two-phonon coupling rates have been demonstrated to reach up to $240$\,Hz \cite{quadratic6,quadratic8,quadratic9} paving the way to phonon shot noise measurements\,\cite{Clerk2010}. Moreover, to prepare nonGaussian quantum states some degree of nonlinearity is necessary so that quadratic coupling plays a fundamental role in many proposed protocols to generate, for example, quantum superpositions\,\cite{super1} and Fock states\,\cite{fock1,fock2}.  Levitation of a nanoparticle in a cavity standing wave is particularly favorable to study quadratic coupling, since there is no external elastic potential. The particle is naturally attracted by the optical gradient force toward an intensity maximum where the coupling is purely quadratic. Here we demonstrate a nonlinear coupling strong enough to cool a levitated nanosphere by more than two orders of magnitude. To the best of our knowledge, this is the first time dominant cooling due to this type of coupling in a cavity is reported. Importantly, the resulting oscillator dynamics is equivalent to that obtained with active parametric feedback\,\cite{Gieseler2012,Gieseler2014}. However, as for the comparison between linear cavity cooling and cold damping\,\cite{vitali2008cold}, the cooling mechanism is passive and does not rely on a position measurement. Finally, we describe the resulting highly nonthermal state of the mechanical motion caused by the nonlinear interaction which compares well with our theoretical description of the experiment.

\begin{figure}[!ht]
\includegraphics[width=8.6cm]{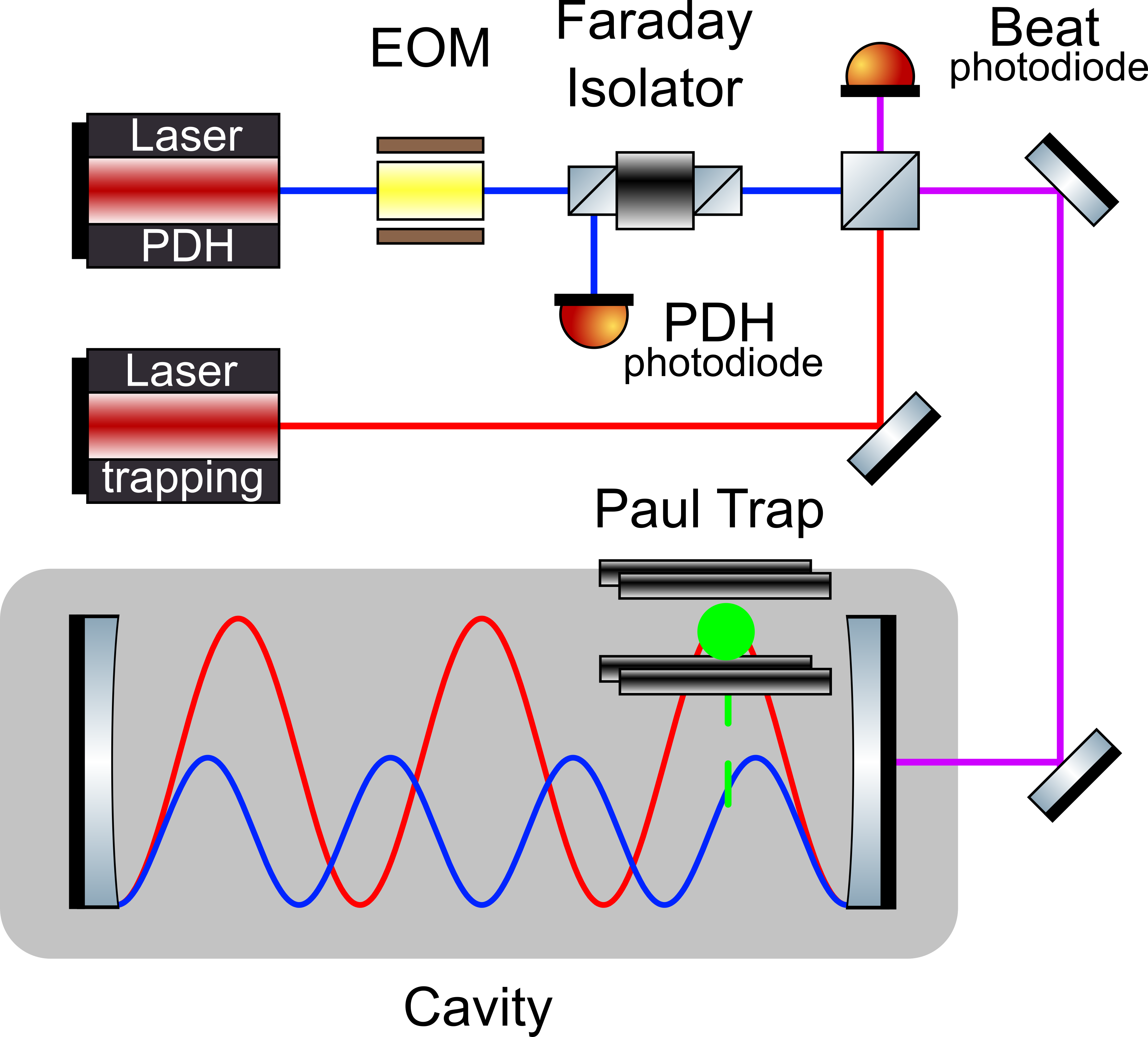}
\caption{Simplified layout of the experiment\,\cite{supp}. A weak probe beam is used to lock the cavity by implementing a PDH scheme. A second beam, generated by a different laser, is used to optically trap the particle. By detecting the beat note of the lasers, the trapping beam is offset phase locked to the PDH beam one FSR apart $\sim\,10.3$\,GHz. The Paul trap, mounted on a three-axis translation stage, is aligned in the cavity transverse direction but kept far from the cavity center. The trapping site can be controlled with a $\sim10\,\mu$m resolution.}
\label{fig1}
\end{figure}

We create our optomechanical system by levitating a highly charged silica nanosphere in a composite potential obtained by overlapping an electrodynamic potential and an optical standing wave. The former provided by a linear Paul trap, the latter resulting from driving a high finesse Fabry-Perot cavity.
In our typical scenario, the nanoparticle is optically trapped along the main axis by the gradient force in one of the cavity antinodes while radial confinement is guaranteed by the Paul trap potential and the transverse beam profile. The presence of a dielectric in the cavity field shifts its resonance frequency by $\Delta(x)=- U_o \,\text{cos}^{2}(kx)$ where $x$ is the particle position along the cavity axis with its origin at the antinode nearest the cavity center, $k=2\pi/\lambda$ the optical field wave number and $U_o=\frac{3}{2}\frac{V}{V_{m}}\frac{\epsilon-1} {\epsilon+2}\,\omega_{l}$, the maximum frequency shift obtained at the antinode of the electric field. Here, $V$ and $V_{m}$ are the sphere and cavity mode volume, respectively, $\epsilon$ denotes the nanosphere permittivity and $\omega_{l}$, the laser frequency. It is clear that $\Delta(x)$ can give rise to a highly nonlinear optomechanical coupling.

We show in Fig.\,\ref{fig1} a schematic overview of our experimental setup. Two Nd:YAG lasers with a wavelength of $\lambda\simeq1064\,$nm drive a $L_{cav}=14.58\pm0.02\,$mm long cavity with a finesse of $\mathcal{F}=36000$ (half linewidth $\kappa/2\pi=143\pm1$\,kHz, input rate $\kappa_{in}/2\pi=69\pm4$\,kHz). The cavity has a nearly confocal configuration with a waist of $w_{st}=62\,\mu$m (i.e. $V_{m}=\pi w_{st}^{2}/4L_{cav}$ ). One laser is exploited as a weak probe field and locked to the cavity by implementing a Pound-Drever-Hall (PDH) scheme. The second is used to optically trap the nanoparticle. Its frequency is offset locked to the weak beam, one Free Spectral Range (FSR$=c/2L_{cav}=10.27\pm0.02\,$GHz) away and its detuning from the cavity resonance can be precisely controlled. Both beams are injected in the cavity with a mode matching $>91$\%.

The particle is charged during the loading process by means of electrospray ionisation and captured directly in medium vacuum in the Paul trap (see Ref.\,\cite{trap,supp}).  We use commercial silica nanospheres of measured mass $m=4.88\pm0.03\times 10^{-17}$\,kg\,\cite{camera} and radius $185\pm2\,$nm. The trap is mounted on a three-axis translation stage. This is important for two reasons. First, it allows us to enhance the linear coupling of the probe field by trapping optically away from the cavity center\,\cite{Kiesel14180}. Second, it allows us to strongly suppress excess micromotion. Indeed, contrary to previous implementations\,\cite{giacomo2015,giacomo2016}, the dynamics along the cavity axis is ideally micromotion free as long as this axis coincides with the main axis of the Paul trap. We can measure and control the position of the optical trapping site referred to the cavity center with a resolution of $\sim10\,\mu$m, mainly limited by the particle thermal variance before optical confinement.

In the following we  focus on the center-of-mass motion (COM) along the cavity axis and assume that the nanoparticle is confined at an antinode of the trapping field. The nonlinear dynamical equation of motion for the oscillator and the optical fields are\,\cite{Monteiro2013}

\begin{equation}\label{eq1}
\begin{split}
\ddot{x}=-\omega_o^{2}x-\frac{\hbar k U_o} {m}\sum_j a_j^\dagger a_j \,\text{sin}[2(k x+\phi_j)]-\gamma_{g}\dot{x}+\frac{\zeta}{m} \\
\dot{a_j}=-(\kappa-i\Delta_o^j) a +i U_o\, a\, \text{cos}^2(k x +\phi_j)+\sqrt{2\kappa_{in}}\alpha_{in,j}+v_j\,.
\end{split}
\end{equation}

\noindent where $j=p,t$ indicate the probe and trap fields, respectively. In Eq.\,\ref{eq1}, $\omega_o$ is the Paul trap secular frequency, $\gamma_{g}$ the gas damping,  $\Delta_o^j$ is the empty cavity detuning, $\kappa=\kappa_{in}+\kappa_{loss}$ is the total cavity half linewidth,  $\alpha_{in,j}$ are the driving amplitudes, $v_i=\sqrt{2 \kappa_{in}}~a_{in,j}+\sqrt{2 \kappa_{loss}}~a_{loss,j}$ is a weighted sum of all vacuum operators. Field fluctuations are uncorrelated with the only nonvanishing correlation function given by $\langle a_i(t) a_j^\dag(t')\rangle = \delta (t-t')\delta_{ij}$ and $\zeta$ is a Brownian stochastic force that arises from background gas collisions and with a correlation function given by$   \langle\zeta(t)\zeta(t'\rangle)= 2\,k_{B}T_{bath}m\,\gamma_{g}\delta(t-t')=S_{th}\delta(t-t')$, where $T_{bath}$ is the temperature of the background gas.

\begin{figure}[!ht]
\includegraphics[width=8.6cm]{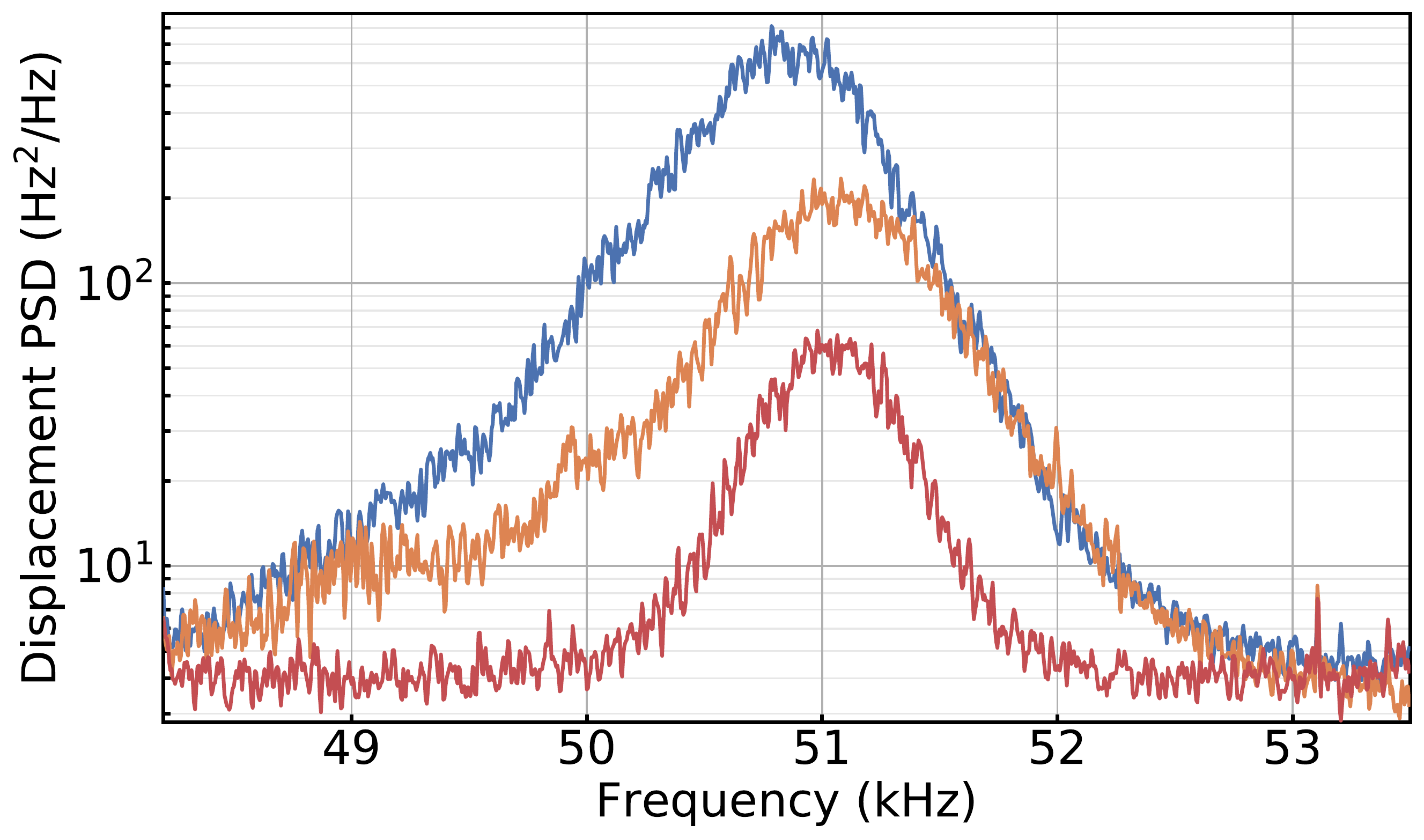}
\caption{Calibrated PSD of the frequency fluctuations induced by the particle motion in the PDH error signal at different pressures. The blue, orange and red PSDs show the particle displacement at pressures $1.2\times10^{-2}\,$mbar, $8.6\times10^{-4}\,$mbar and $5.4\times10^{-6}\,$mbar, respectively.}
\label{fig2}
\end{figure}

If we consider a scenario where the probe power is significantly weaker than the trap beam, i.e. $\alpha_{in,p}\ll\alpha_{in,t}$, it is quite immediate to see that upon usual expansion around the steady state solution to second order of the trigonometric functions in Eq.\,\ref{eq1} one finds a purely quadratic coupling for the trapping beam with $G_{2,t}=k^2U_o$ (i.e., $\phi_t=0$), while the probe field has both a linear and quadratic coupling respectively given by $G_{1,p}=k U_o\,\text{sin}(2\phi_p)$ and  $G_{2,p}=k^2 U_o\,\text{cos}(2\phi_p)$. In these last two expressions the phase is entirely determined by the position $x_o$ of the localisation site referred to the cavity center, i.e., $\phi_p= \pi/2+\pi x_o/L_{cav}$. To gather a clearer understanding of the oscillator dynamics, however, it is more convenient to trace out the cavity rather that linearize, i.e., we want to write an approximate equation of motion for the particle dynamics in the following form

\begin{equation}\label{eq2}
  \ddot{x}=-\Omega_m^2 x\,\left(1+\epsilon_D x^2\right)- \left(\gamma_{g}-\Omega_m^2 \gamma_{nl} x^2\right)\frac{p}{m}+\frac{\zeta}{m}
\end{equation}

\begin{figure*}[!ht]
\includegraphics[width=1\textwidth]{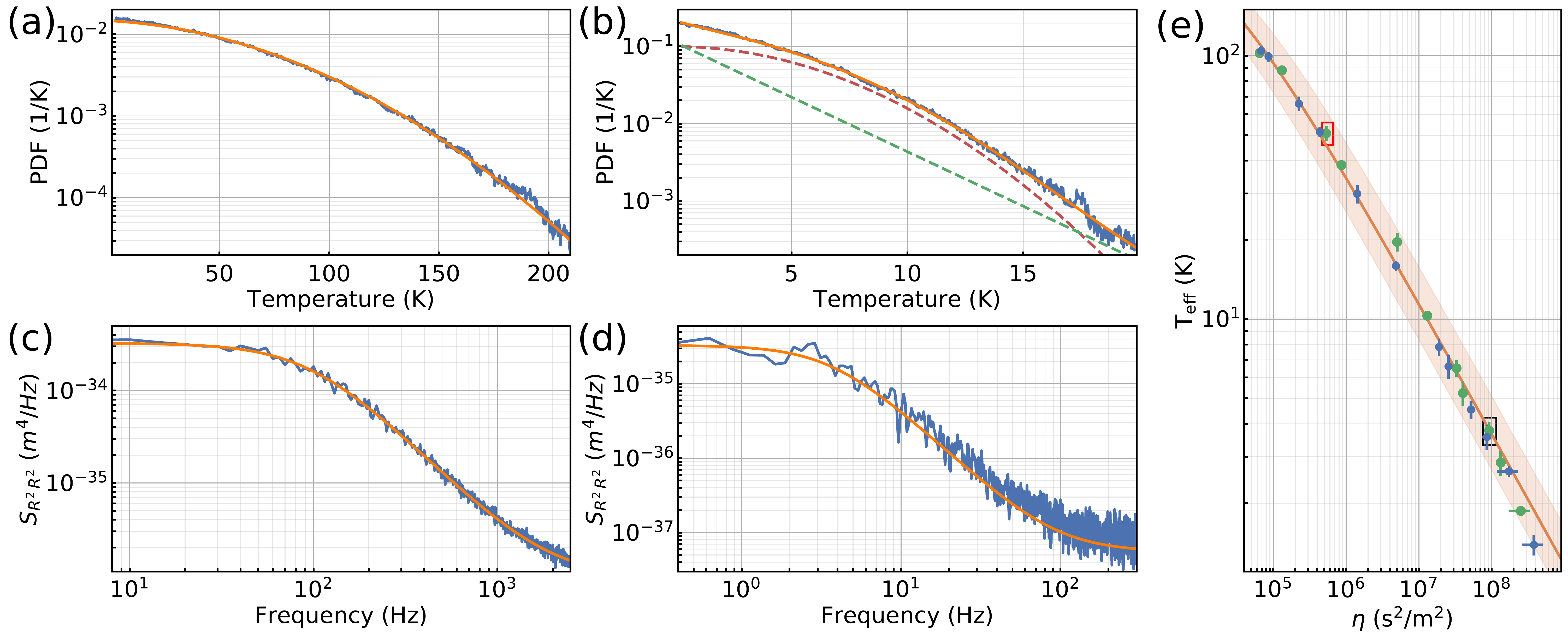}
\caption{Panels (a) and (b): Energy distribution (blue), expressed in units of $k_B$, for the motion along the \textit{x}-axis along with a fit (orange) following Eq.\,\ref{eq5} for different pressures $P_g=8.6 \times 10^{-2}$\,mbar and $1.6\times10^{-4}$\,mbar respectively. In Panel (b) the contribution of the detection noise is included; the corresponding marginal distributions for the motion (red dashed) and noise (green dashed) are also shown. Panels (c) and (d) show the PSD of $R^2$ corresponding to the distributions in panels (a) and (b) respectively; fits (orange) of the $R^2$ spectra allow a direct estimation of the energy autocorrelation time constant (see main text); Panel (e): effective temperature as function of fitted parametric gain $\eta=\gamma_{nl}/\gamma_g$ along with \textit{a priori} analytical estimation (orange line). Shaded region indicates the uncertainty of the theoretical curves due to experimental uncertainty in relevant parameters; blue and  green data points correspond to localization on slightly different optical wells. The red square marks the datapoint corresponding to panels (a) and (c), the black square to panels (b) and (d).
}
\label{fig3}
\end{figure*}

\noindent where $\Omega_m$ is the optical trap frequency and where we have introduced two additional terms: an elastic Duffing nonlinearity $\epsilon_D$ and a Van der Pol nonlinear damping $\gamma_{nl}$. The latter corresponds to a dissipation process that becomes more efficient for large amplitude oscillations. Eqs.\,\ref{eq1} can be rewritten in the form of Eq.\,\ref{eq2} by looking at a first order correction to the adiabatic approximation. Neglecting the effect of the probe beam, assuming that $\Omega_m/\kappa\ll 1$, and by following the method described in Ref.\,\cite{elimination}, one finds that

\begin{equation}\label{eq3}
\epsilon_D=\frac{2G_2}{\kappa}\frac{\delta}{(1+\delta^2)},\,\,\,\,
\gamma_{nl}=\frac{8G_2}{\kappa^2}\frac{\delta}{(1+\delta^2)^2},
\end{equation}

\noindent where $\delta$ is the normalized hot cavity detuning of the trapping beam and $\Omega_m^2 = \frac{2 \hbar G_2}{m} |\alpha_{t,s}|^2$ as expected, with $\alpha_{t,s}$ the steady state intracavity field amplitudes. Eqs.\,\ref{eq3} are valid under the additional condition $G_2 <x^2>\ll\kappa$ which is always satisfied in our experiment. As for the linear coupling the oscillator dynamic depends critically on the detuning sign. For a red detuned ($\delta<0$) trapping beam the optical potential is softened and dissipation increased while the opposite happens for a blue detuned beam, which can result in dynamical instability. Interestingly, both nonlinear coefficients are power independent, this is a characteristic inherently due to levitation since there is no intrinsic elastic potential.

It is quite convenient, at this point, to move to a reference frame rotating at $\Omega_m$ and to write the equation of motion for the amplitude $R(t)$ and phase $\varphi(t)$ of the oscillator. By performing deterministic and stochastic averaging\,\cite{strato1967,Roberts1986,Boujo2017}, valid in the high Q limit, one obtains two first order differential equations

\begin{equation}\label{eq4}
\begin{split}
   \dot{R} &= -\frac{\gamma_g}{2} R + \frac{\Omega_m^2 \gamma_{nl}}{8} R^3+\frac{S_{th}}{4m^2\Omega_m^2 R}+\xi =-\frac{d\mathcal{V}(R)}{d R}+\xi\\
   \dot{\varphi}  &= \frac{3\Omega_m}{8}\epsilon_D R^2+\frac{1}{R} \chi.
\end{split}
\end{equation}

\noindent Here, $\xi$ and $\chi$ are two uncorrelated stochastic variables with correlation function $\langle\xi(t)\xi(t'\rangle)=\langle\chi(t)\chi(t'\rangle)= (S_{th}/2m^2\Omega_m^2)\delta(t-t')$ and where we introduced the potential $\mathcal{V}(R)$. Eqs.\,\ref{eq4} allows us to highlight two key aspects. First, the effect of the Duffing term is relegated to the evolution of the phase and has no effect on the energy of the oscillator. Second, the evolution of the amplitude is phase independent. The latter is of particular importance, since it allows us to write a simple one dimensional Fokker-Planck (FP) equation for the evolution of the probability density function (PDF) of R whose steady state solution is well known and given by $P_{\infty}(R)=\mathcal{N}\text{exp}(-4m^2\Omega_m^2 \mathcal{V}(R)/S_{th})$. This, reduces to the Rayleigh distribution in the limit of vanishing nonlinear damping. Expressing the steady state solution of the FP equation in terms of energy $E$ rather than amplitude, one readily finds

\begin{equation}\label{eq5}
  P_{\infty}(E)=\frac{\mathcal{N}}{m \Omega_m^2}\text{exp}\left[-\frac{E}{k_{B} T_{bath}}\left(1+ \frac{\gamma_{nl}}{4 m \gamma_g }E\right)\right]
\end{equation}

\noindent where $\mathcal{N}$ is a normalization constant such that $\int_{0}^{\infty}P_{\infty}(E)dE=1$. Since the energy distribution is known, all the relevant dynamical parameters can be obtained, e.g., the effective temperature and damping. For a vanishing nonlinear damping Eq.\,\ref{eq5} becomes the usual Boltzmann-Gibbs distribution.

It is important to notice that the coupled dynamics described here is formally equivalent to active parametric feedback\,\cite{Gieseler2014}, indeed, Eq.\,\ref{eq5} describes the steady state energy distribution of both processes. This should not come as a surprise since in both cases the oscillator dynamic is modified by an optical force proportional to $x^2p$, i.e., a Van der Pol nonlinear damping. As such quadratic coupling can also be viewed as passive parametric feedback. Furthermore, Eq.\,\ref{eq5} represents the classical limit of two phonon cooling in the quantum regime\,\cite{Nunnenkamp}.

Here, we present data obtained with a probe and trap beam input power of $2.9\,\mu$W and $830\,\mu$W respectively. The probe is locked near resonance while the trap beam has a nominal red detuning of $\Delta_t/2\pi\simeq-100\,$kHz. The particle motion is monitored through the PDH error signal since its linear coupling enables us to measure the mechanical motion directly. Spectra at different pressures are shown in Fig.\,\ref{fig2}. It is clear that the oscillator resonance at $\Omega_m/2\pi\simeq51\,$kHz does not converge to a Lorentzian-like peak, as the pressure is reduced, but rather converges to a Gaussian peak. This apparent broadening is due to low frequency intensity fluctuations and by crosscoupling with the motion in the directions perpendicular to the cavity axis. This motion is adiabatically eliminated when moving to the rotating frame as is the case for the Duffing nonlinearity.

The experimental energy distribution can be obtained from the square of the oscillator amplitude $R(t)$\,\cite{linewidth}. This is shown in Fig.\,\ref{fig3}(a) and (b) at two different pressures where the distributions are expressed in units of $k_B$, i.e. temperature, for a more intuitive reading. At the higher pressure of $P_g=8.6 \times 10^{-2}$\,mbar (panel (a)), the deviation from a thermal exponential distribution is immediately recognizable, indeed, the nonlinear damping is much more efficient in suppressing large amplitude fluctuations. At the lower pressure of $P_g=1.6\times10^{-4}$\,mbar (panel (b)) this behaviour is initially less evident. As the motion becomes colder the impact of the detection noise becomes more relevant and it needs to be taken into account. Assuming the noise floor is white and uncorrelated with the motion, its distribution is again exponential. We fit the experimental data taking into account both processes, the oscillator energy PDF is then recovered by taking the marginal distribution. As the intrinsic gas damping cannot be measured independently, and pressure gauges have a rather low accuracy, we use as fitting parameter the ratio $\eta=\gamma_{nl}/\gamma_g$ which can be interpreted as a parametric gain.

We show in Fig.\,\ref{fig3} (panel (c)) the experimental effective temperature $T_{eff}$ as a function of $\eta$, along with an \textit{a priori} analytical estimation. The final temperature can be estimated in two ways: from the area of the peak in the probe PSD, as is typical, and from the expectation value of the fitted distribution. The consistency of these two estimates, well within the experimental uncertainty, and the agreement with the analytical expectation, demonstrate that even at the lowest pressure the quadratic coupling is the dominating process in the dynamics.

\begin{figure}[!ht]
\includegraphics[width=8.6cm]{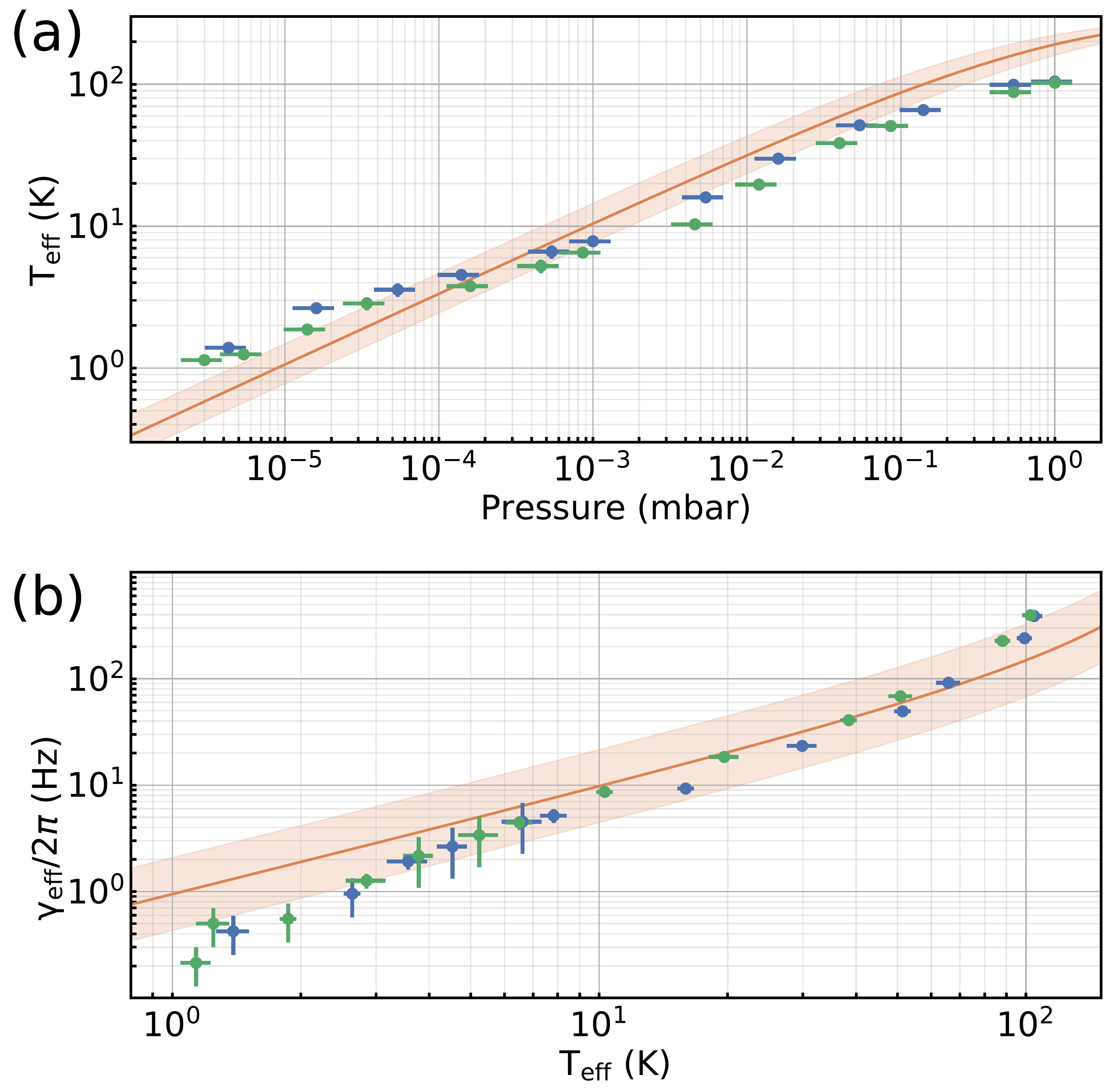}
\caption{Panel (a): The effective temperature as a function of pressure. Panel (b): The measured effective damping as a function of $T_{eff}$. In both panels blue and  green data points correspond to localization on slightly different optical wells; also shown is an analytical estimation (orange line) with the shaded region indicating the uncertainty of the theoretical curves.
}
\label{fig4}
\end{figure}

Another clear signature of the nonlinear damping, is the dependance on pressure of the effective temperature. An approximate expression can be obtained from Eq.\,\ref{eq5}; in the low pressure limit, we have $T_{eff}=(4 m T_{bath}/\pi k_B \eta)^{1/2}$. Since $\eta$ grows inversely proportionally to the pressure, the effective temperature decreases proportionally to the square root of the pressure. This is shown in Fig.\,\ref{fig4} where the experimental observation is compared with analytical estimation. A direct consequence is that the effective total damping $\gamma_{eff}$ must have a similar behaviour; in the low pressure limit we have $\gamma_{eff}=\gamma_g\,(\pi k_B T_{bath} \eta/4 m)^{1/2}$. This implies perfect correlation between $T_{eff}$ and $\gamma_{eff}$ as the pressure is reduced.

A direct estimation of the effective damping can be obtained even in the presence of the Duffing term and of the broadening of the spectral peak, evident in Fig.\,\ref{fig2}. By looking at the PSD of $R^2(t)$ it is possible to obtain information on the energy autocorrelation time constant. PSDs at two different pressures are shown in Fig.\,\ref{fig3}. Although possible for a thermal oscillator\,\cite{linewidth}, calculating an analytical expression for the $R^2$ PSD when the dynamic is dominated by a nonlinear damping is not trivial. However, it can be shown that modeling the PSD as $S_{R^2R^2}(\omega)=16 \gamma_R a_o^2/(\omega^2+\gamma_R^2)$ then $\gamma_R$ allows us to calculate the effective damping as $\gamma_{eff}=\gamma_R \,\sigma_E^2/\langle E\rangle^2$, where $\langle E\rangle^2$ and $\sigma_E^2$ are the energy mean and variance respectively, both of which can be calculated from the experimental distribution. The effective damping calculated with this method plotted as a function of $T_{eff}$ is shown in Fig.\,\ref{fig4}(b) demonstrating the expected good correlation.

In conclusion, we have demonstrated strong quadratic cooling of the center-of-mass motion of a nanoparticle optically trapped in a cavity standing wave.  Comparison of the experimental results show good agreement with analytical \textit{a priori} model. This type of passive parametric feedback cooling is analogous to the active feedback cooling in levitated optical tweezers and indeed comparable temperatures are obtained here. The major difference between the two methods is that the cavity automatically applies feedback whereas conventionally detection and electronic feedback are required to modulate the potential for active cooling.  We have also shown that a highly nonthermal state is produced, which in the quantum regime, would be a highly nonclassical state. In both cases, the cooling rate decreases as the particle is cooled to the bottom of the optical potential. Among the main advantages of optical levitation in a cavity field lies in the possibility of manipulating the coupling from quadratic to linear and vice versa by simply controlling the power ratio between the probe and trapping field. As linear coupling is significantly more efficient than quadratic cooling it should be possible to switch between the two configurations producing two nonequilibrium steady states at different effective temperatures. This is a new tool for exploring nanoscale thermodynamics allowing the measurement of the relative entropy change for testing nonequilbrium thermodynamics and fluctuation theorems\,\cite{ness1,ness2,ness3}. An even more intriguing possibility is to exploit a similar protocol where the particle is initialized close to the quantum ground state\,\cite{Delic2020science} through linear coupling and then to adiabatically change to a quadratic coupling. Such a scheme may allow squeezing of mechanical motion\,\cite{Nunnenkamp} but also the creation of other nonclassical states by using the nonlinear damping demonstrated here.

The authors would like to acknowledge useful discussions with M. Toro\v{s}. The authors acknowledge funding from the EPSRC Grant No. EP/N031105/1. N.P.B. acknowledges funding from the EPSRC Grant No. EP/L015242/1. A.P. has received funding from the European Union’s Horizon 2020 research and innovation programme under the Marie Sklodowska-Curie Grant Agreement No. 749709.

\bibliographystyle{nicebib}
\bibliography{gtwo_bib}

\end{document}